# Peak separation methods for inverse photoelectron spectra: Comparing second derivative, curve fitting, and deconvolution analyses

Ryotaro Nakazawa[1,a), b)], Haruki Sato[1], and Hiroyuki Yoshida[1, 2, , b)]

[1]Graduate School of Engineering, Chiba University, Chiba 263-8522, Japan
[2]Molecular Chirality Research Center, Chiba University, Chiba 263-8522, Japan

**Abstract**

Inverse photoelectron spectroscopy (IPES) is a powerful technique for probing the unoccupied electronic states of materials. It can be regarded as the inversion process of photoelectron spectroscopy (PES), which examines the occupied states. Recently developed low-energy inverse photoelectron spectroscopy (LEIPS) can significantly advance the study of unoccupied states, owing to an improved signal-to-noise ratio and minimal sample damage compared to conventional IPES. However, the instrumental resolution remains at 0.2 eV, which is one order of magnitude lower than that of PES. Spectral broadening caused by the low instrumental resolution often results in overlapping peaks. Peak separation is therefore crucial in the analysis of LEIPS spectra. In this study, we compared three peak separation methods: second derivative, curve fitting, and deconvolution. These methods were applied to modeled and experimental LEIPS spectra of the lowest unoccupied molecular orbital-derived band of pentacene, which consists of two splitting peaks due to the two inequivalent molecules in the unit cell. We systematically and quantitatively evaluated the performance of each method in terms of analysis parameters and discussed its robustness to noise as well as its peak separation capability. This work offers a practical framework for peak separation in LEIPS, with extensions to PES and a wide range of spectroscopies.

[a)] Present address: Institute for Molecular Science, Okazaki 444-8585, Japan
[b)] Authors to whom correspondence should be addressed: nakazawa@ims.ac.jp; hyoshida@chiba-u.jp

# I. *Introduction*

Electron spectroscopy is a powerful tool for studying the electronic states of materials. X-ray photoelectron spectroscopy (XPS) examines core levels and provides information on constituent elements and chemical states. Ultraviolet photoelectron spectroscopy (UPS) and inverse photoelectron spectroscopy (IPES) furnish information on the energy and density of occupied and unoccupied states, respectively. Unfortunately, peaks in the electron spectra are often not well separated because of the intrinsic broadening of the material's electronic states, the photoemission process (*e.g.*, lifetime broadening), and instrumental resolution. Various peak separation methods have been applied. For example, XPS spectral features are often separated into several peaks by fitting with Gaussian, Lorentzian, or Voigt functions. In UPS, particularly in angle-resolved UPS (ARUPS), second derivative[1, 2] or curve fitting[3–6] is the standard method for determining peak positions.

In some cases, closely spaced peaks are not resolved due to low instrumental resolution. Because the observed spectrum is a convolution of the original (or true) spectrum and the instrumental function, peaks can be restored by deconvoluting the observed spectrum with the instrumental function. To date, deconvolution has been applied to PES[7,8] and electron energy loss spectroscopy[9], improving energy resolution by a factor of two. In principle, deconvolution should also be useful for IPES spectra because the energy resolution of IPES typically ranges from 0.2 eV[10] to 0.5 eV[11] Peak separation capability is mainly limited by instrumental resolution. However, a high signal-to-noise (S/N) ratio is necessary for reliable deconvolution calculations. As the cross-section of IPES is three to five orders of magnitude smaller than that of PES[1], the S/N ratio of IPES is generally very low.

We have developed low-energy inverse photoelectron spectroscopy (LEIPS)[12–14], in which the kinetic energy of the electron is reduced to the damage threshold of organic materials (typically 5 eV), thus suppressing sample damage [15]. Because of energy conservation, the emitted photons are in the near-ultraviolet range, enabling the bandpass filter and the photomultiplier tube to be used for photon detection, improving energy resolution down to 0.25 eV and the S/N ratio[13]. The high S/N ratio of LEIPS should allow the use of deconvolution for peak separation.

Recently, we developed angle-resolved LEIPS (AR-LEIPS)[16], which revealed the conduction band electronic structures of organic materials[17] and solar-cell perovskite materials[18]. Separating peaks and determining peak positions were critical in analyzing the AR-LEIPS spectrum because the subtle angle-dependent energy position is important to AR-LEIPS. While we tentatively applied the second derivative method in the previous work, the choice of method and its conditions require further investigation.

In this study, we quantitatively and systematically compare three peak separation methods—second derivative, peak fitting, and deconvolution—and provide practical guidelines for separating overlapping peaks in LEIPS spectra. The second derivative and the peak fitting methods are commonly used for UPS and XPS analyses. Conversely, there are few studies utilizing the deconvolution method for analyzing IPES spectra[19,20]. We demonstrate the applicability of deconvolution to LEIPS spectra using nonlinear iterative deconvolution algorithms.

We analyze the LUMO-derived spectrum of pentacene as an example, as we recently investigated the band split in the AR-LEIPS spectra[17]. Because the pentacene crystal contains two molecules in the unit cell, the LUMO band splits into upper and lower bands with an energy difference of 0.4 eV at the Γ point[17]. To benchmark the peak separation methods, we also analyze the model spectrum composed of two overlapping peaks. The peak intensities, the energy difference, and the peak widths are chosen to be similar to the LUMO of the pentacene film.

## II. *Methods*

### A. Second derivative

The second derivative is the most commonly used method for determining peak positions in ARUPS. This method highlights the positions of overlapping peaks. As derivation essentially consists of calculating the difference between the adjacent intensities divided by the energy interval, small noises are amplified. Consequently, the measured spectrum must be smoothed before operation. However, the smoothing parameter, such as the smoothing width, introduces arbitrariness into the results; insufficient smoothing introduces artificial structures, whereas excessive smoothing suppresses spectral structures.

In this study, the Savitzky-Golay (SG) method is used for smoothing. In the SG method, a specific point and its neighbors are fitted to an $n$th-order polynomial function, and the intensity of the specific point is replaced by that of the $n$th-order polynomial function. Consequently, the uncertainties of the peak energies determined by the second derivative are: (i) the degree of the polynomial function, (ii) the number of iterations, and (iii) the number of smoothing points (smoothing window size). In this study, we use a second-order polynomial function and treat the number of smoothing points as a free parameter. The iteration number is fixed at four because peak energies are reliably determined when the SG method is performed three or more times. Further increasing the iteration number makes the results more sensitive to the number of smoothing points [see Fig. S1 in Supplementary Materials].

### B. Curve fitting

Curve fitting is probably the most widely used method for peak separation in XPS. A linear combination of functions (Gaussian, Lorentzian, or Voigt function) is fitted to an experimental spectrum using the nonlinear least squares method. The result is dependent on the initial conditions of the nonlinear least squares method. Often, we do not know the choice and number of fitting functions. An insufficient number of functions cannot correctly reproduce the shape of the experimental spectral line. On the other hand, too many functions impose a mutual dependence of the fitting parameters. To avoid this, we often impose restrictions on peak width or intensity to obtain reasonable results. A systematic study of peak fitting conditions is beyond the scope of this study.

To separate overlapping peaks from a measured spectrum, the measured spectrum is fitted with a function. In the fitting procedure, the measured spectrum $m(x)$ as a function of energy ($x$) is fitted with $f(\beta, x)$, as $m(x) = f(\beta, x) + r(x)$, where $\beta$ is the parameter of the fitting function and $r(x)$ is the residual. The optimal $\beta$ is obtained by the nonlinear least squares method, which minimizes the sum of squares error, $S(\beta) = \Sigma[f(\beta, x) - m(x)]^2$. In this work, we use the Levenberg-Marquardt algorithm for the nonlinear least squares method without parameter constraints[21].

### C. Deconvolution

Deconvolution restores an original spectrum $o(x)$ by removing the noise component and broadening by instrumental function $i(x)$ from a measured spectrum $m(x)$. Here, $m(x)$ is expressed as follows:

$$m^*(x) = o(x) \otimes i(x) + N(x) \quad (1)$$

$$m(x) = P[m^*(x)] \quad (2)$$

where $\otimes$ refers to the convolution operation, $N(x)$ means the background of $m(x)$, and $P(\lambda)$ is the Poisson distribution of the mean $\lambda$. A measured spectrum without noise $m^*(x)$ is expressed as Eq. (1). As the intensity of each measured spectral point means a count of random events that have a constant average value, each spectral intensity of $m(x)$ has Poisson-distributed fluctuations around $m^*(x)$.

In the deconvolution method, first, $N$ is subtracted from $m(x)$, and $m(x)$ and $i(x)$ are smoothed by auto/cross-correlation prefiltering[22], that is, $i(x) \otimes i(-x) = f(x)$ and $m(x) \otimes i(-x) = g(x)$, then $f(x) \rightarrow i(x)$ and $g(x) \rightarrow m(x)$. $m(x)$ is deconvoluted with $i(x)$ using nonlinear iterative deconvolution algorithms[23]. In the main text, we focus on Jansson's method[24] because this method gives more precise results than van Cittert's method with positive constraints[25] and is more robust to noise than Gold's ratio method[26]. Other methods are treated in Supplementary Material (Figs. S2 and S3, and Table S1). In Jansson's method, the following recursion equation is calculated,

$$\hat{o}^{k+1}(x) = \hat{o}^k(x) + r[\hat{o}^k(x)][m_0(x) - i(x) \otimes \hat{o}^k(x)] \quad (3)$$

$$r[\hat{o}^k(x)] = r_0 \left[1 - \frac{2}{a-b}\left|\hat{o}^k(x) - \frac{a+b}{2}\right|\right]. \quad (4)$$

Here, $\hat{o}^k(x)$ is an estimated original spectrum calculated by the above equations $k$ times, namely, a deconvoluted spectrum. $r_0$, $a$, and $b$ are constants. At $k = 0$, $m(x)$ is set as $\hat{o}^0(x)$. In the main text, we will treat the iteration number as the free parameter to be optimized. The treatment of $r_0$, $a$, and $b$ is described in Supplementary Material (Figs. S4–S7).

## III.  *Data preparation*

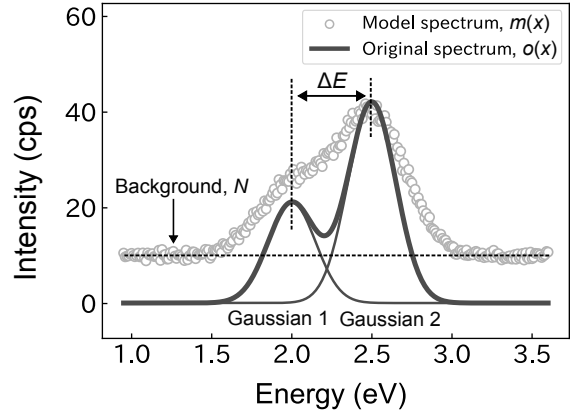

**FIG. 1. Construction of model spectrum *m(x)*. The measured spectrum is modeled by convolving the original spectrum *o(x)* with the instrumental function *i(x)*, adding the background *N*, and introducing noise with a Poisson distribution. *o(x)* is the linear combination of Gaussian functions called Gaussians 1 and 2. The peak intensity ratio of Gaussians 1 and 2 is 1:2, and the energy spacing of these peaks Δ*E* is 0.50 eV.**

First, we analyze the model spectrum $m(x)$. Because we already know the original spectrum $o(x)$, analyzing the model spectrum provides us with both peak separation capability and precision. $o(x)$ is constructed to reproduce the experimental LEIPS spectrum of pentacene. In Fig. 1, $o(x)$ is composed of two Gaussian functions called Gaussians 1 and 2. The energy spacing of these peaks $\Delta E$ is 0.5 eV, and the full width at half maximum (FWHM) of Gaussians 1 and 2 in $o(x)$ is 0.35 eV. The peak intensity ratio is 1:2. The model spectrum $m(x)$ is obtained by convoluting $o(x)$ with the instrumental function $i(x)$. $i(x)$ is assumed to be a Gaussian function with an FWHM of 0.35 eV, determined experimentally from the Fermi edge of a LEIPS spectrum of a polycrystal Ag surface. The background $N(x)$ is assumed to be constant, and its intensity is set to 1/4 of the peak intensity of $o(x)$. Using Eqs. (1) and (2), $m(x)$ is constructed. Because we assume that the noise follows the Poisson distribution of the noiseless model spectrum $m^*(x)$, the standard deviation of each $m(x)$ point is $\pm\sqrt{m^*(x)}$. The S/N ratio is changed by varying the intensity of $m^*(x)$. We introduce the root mean square error (RMSE) $E_{\text{RMS}}^{\text{m}}$ between $m(x)$ and $m^*(x)$ to quantify the noise,

$$E_{\text{RMS}}^{\text{m}} = \sqrt{\frac{1}{n}\sum[m(x) - m^*(x)]^2} \qquad (5)$$

where $n$ is the number of data points in $m(x)$. $E_{\text{RMS}}^{\text{m}}$ of $m(x)$ used in Fig. 1 is 0.73.

In Sec. V A, we discuss the robustness to noise of the three methods. To this end, we prepare a series of $m(x)$ with different noise levels with $E_{\text{RMS}}^{\text{m}}$ ranging from 9.44 to 0.78. At first, the background $N(x)$ is multiplied by 8, and we generate a spectrum with $E_{\text{RMS}}^{\text{m}} = 9.44$. Then, by averaging the spectra generated by the same method, we obtain $m(x)$ with a smaller $E_{\text{RMS}}^{\text{m}}$. In Sec. V B, we discuss peak separation capability. The energy spacing $\Delta E$ of Gaussians 1 and 2 is changed from 0.45 eV to 0.15 eV in steps of 0.05 eV. We set the FWHMs of Gaussians 1 and 2 in $o(x)$ at 0.20 eV. Because the analysis results are affected by noise, $E_{\text{RMS}}^{\text{m}}$ of $m(x)$ is set at approximately 0.3.

Second, we analyze the experimental LEIPS data. Pentacene is deposited to a Cu(110) surface at a thickness of 10 nm by vacuum vapor deposition. LEIPS measurement is performed at 90° to the surface at a sample temperature of -55°C. The central wavelength of the detected photons is 257 nm.

In the realistic LEIPS measurements, as the noiseless LEIPS spectrum [$m^*(x)$] is unknown, we define the RMSE value between the LEIPS spectrum $m(x)$ and smoothed LEIPS spectrum $m_{\text{SG}}(x)$, $E_{\text{RMS}}^{\text{s}}$, as a measure of the noise quantity instead of $E_{\text{RMS}}^{\text{m}}$,

$$E_{\text{RMS}}^{\text{s}} = \sqrt{\frac{1}{n}\sum[m(x) - m_{\text{SG}}(x)]^2} \qquad (6)$$

where $n$ is the number of data points. $m_{\text{SG}}(x)$ is obtained by the SG method with reasonable parameters (see Sec. IV). In Sec. V A, we discuss the robustness to noise of the methods for the experimental LEIPS spectra taken from the literature.[16] These spectra are systematically measured as a function of the total incident charge of the irradiated electron beam (integration time in the measurements), resulting in a different S/N ratio.

# IV.  Results

## A. Analysis of model spectra

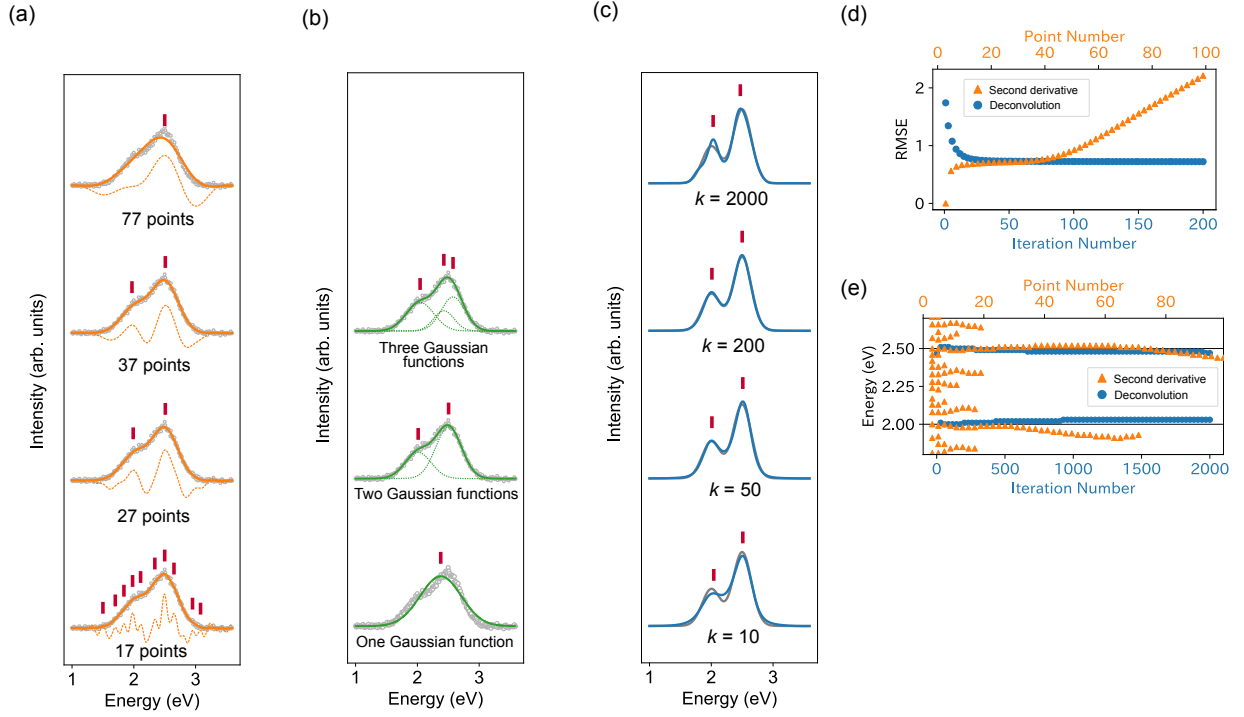

**FIG. 2**. Peak separation analysis of the model spectrum $m(x)$ (gray circles) shown in Fig. 1. Red bars indicate the obtained peak positions, and $o(x)$ is shown as a gray solid line.
(a) Second derivative analysis. The Savitzky–Golay (SG) method is used for smoothing. Smoothed spectra with various numbers of smoothing points (orange solid lines) and their negative second derivative spectra (orange dashed lines). (b) Curve fitting analysis. The fitted spectrum (green solid line) and the composed Gaussian functions (green dashed lines). (c) Deconvolution analysis. Deconvoluted spectra with various iteration numbers $k$ (blue solid lines). (d) Root mean square error (RMSE) of the deconvoluted spectrum ($E_{RMS}^d$, blue plots) as a function of $k$, and RMSE of the smoothed spectrum ($E_{RMS}^s$, orange plots) as a function of the number of smoothing points. (e) Peak energies determined by the second derivative (orange) and deconvolution (blue). Black solid lines indicate the true peak energy of $o(x)$.

  The model spectrum $m(x)$ shown in Fig. 1 is analyzed to evaluate the peak separation capabilities of the second derivative, curve fitting, and deconvolution methods. First, we perform the second derivative analysis to obtain the peak energies of Gaussians 1 and 2. Figure 2(a) shows the model spectrum, the smoothed spectra, and the negative second derivative spectra. Because $m(x)$ is smoothed before the second derivative, the results depend on the number of smoothing points. When 17 points are used, the negative second derivative spectrum has many extra peaks due to insufficient smoothing. When 27 and 37 points are used, these additional peaks disappear, and two peaks are observed. When 77 points are used, the smoothed spectrum is broader than $m(x)$, and only a single peak is obtained.

  To provide a criterion for selecting a reasonable number of smoothing points, we analyze the noise level $E_{RMS}^s$ as a function of the number of smoothing points in Fig. 2(d). $E_{RMS}^s$ increases rapidly from zero to five smoothing points because $E_{RMS}^s$ is zero before the smoothing. Then, $E_{RMS}^s$ is constant from five to 35 smoothing points. In this region, the smoothed spectrum preserves the line shape of $o(x)$. Beyond 37 points, $E_{RMS}^s$ increases again, indicating excessive smoothing. Therefore, the upper limit must be 35 points. Next, we examine the number of peaks as a function of smoothing points [Fig. 2(e)] to determine the lower limit. Because two peaks are obtained when the number of smoothing points is $\geq 21$, the lower limit must be 21 points. The optimum number of smoothing points ranges from 21 to 35, and the obtained peak energies are 1.97–1.99 and 2.50–2.52 eV in this range. The energy variation in this region can provide a measure of the uncertainties. If a reasonable number of peaks is not obtained before $E_{RMS}^s$ increases, that may signify that the statistics of the original data are not sufficient (see Sec. V A).

Figure 2(b) shows the results of curve fitting. We fit the model spectrum with linear combinations of one to three Gaussian functions. The model spectrum cannot be fitted with a single Gaussian function. The fit with the two Gaussian functions reasonably reproduces the model spectrum. The peak energies of the fitted spectrum are 2.01 and 2.51 eV, in good agreement with the original peak positions with an error of 0.01 eV. The three Gaussian functions also reproduce $m(x)$ well. From this analysis alone, we cannot determine the number of Gaussian functions involved in the model spectrum. Analysis of peak separation by curve fitting requires careful selection of the fitting function and the number of functions.

Finally, we perform the deconvolution. We find that the number of iterations $k$ greatly affects the deconvoluted spectrum $\hat{o}^k(x)$. Figure 2(c) shows the deconvolution results. At $k = 10$, the line shape of $\hat{o}^k(x)$ is broader than that of the original spectrum $o(x)$. At $k = 20$ and 200, the line shape of $\hat{o}^k(x)$ is consistent with that of $o(x)$. At $k = 2000$, an additional shoulder around 1.8 eV appears. We should find the optimum $k$ to reliably restore $o(x)$.

To establish the criterion to optimize $k$, we introduce the RMSE between the model spectrum $m(x)$ and the convolution of $\hat{o}^k(x)$ and $i(x)$ [$\hat{m}^k(x)$], $E_{RMS}^d$, as a measure of difference between the original and deconvoluted spectra,

$$E_{RMS}^d = \sqrt{\frac{1}{n}\sum[m(x) - \hat{m}^k(x)]^2} \qquad (7)$$

where $n$ is the number of data points in $m(x)$. Figure 2(d) shows the variation of $E_{RMS}^d$ as a function of $k$. $E_{RMS}^d$ decreases rapidly until $k = 50$ and then remains almost unchanged. When iterative deconvolution is excessively performed (*e.g.*, $k = 2000$), the line shape of $\hat{o}^{2000}(x)$ is distorted although the $E_{RMS}^d$ value remains almost unchanged (not shown). Because $E_{RMS}^d$ is calculated using $\hat{m}^k(x) = \hat{o}^k(x) \otimes i(x)$, the convolution with $i(x)$ obscures the fine distortions of $\hat{o}^{2000}(x)$. To avoid this distortion, we should stop the iteration when the $E_{RMS}^d$ value converges. In this case, we should terminate the iteration at approximately $k = 50$.

We examine the peak energy dependence of $\hat{o}^k(x)$ on $k$ [Fig. 2(e)]. When the $E_{RMS}^d$ value converges (*e.g.*, $k = 50$ and 200), the peak energies agree with the original value with an error of 0.01 eV. In the region of excessive iteration (*e.g.*, $k = 2000$), the peak energy is consistent with that of $o(x)$ with an error of 0.03 eV. The deconvolution determines the peak energies with an uncertainty of less than 0.01 eV after convergence and before excessive processing.

## B. Analysis of experimental spectrum

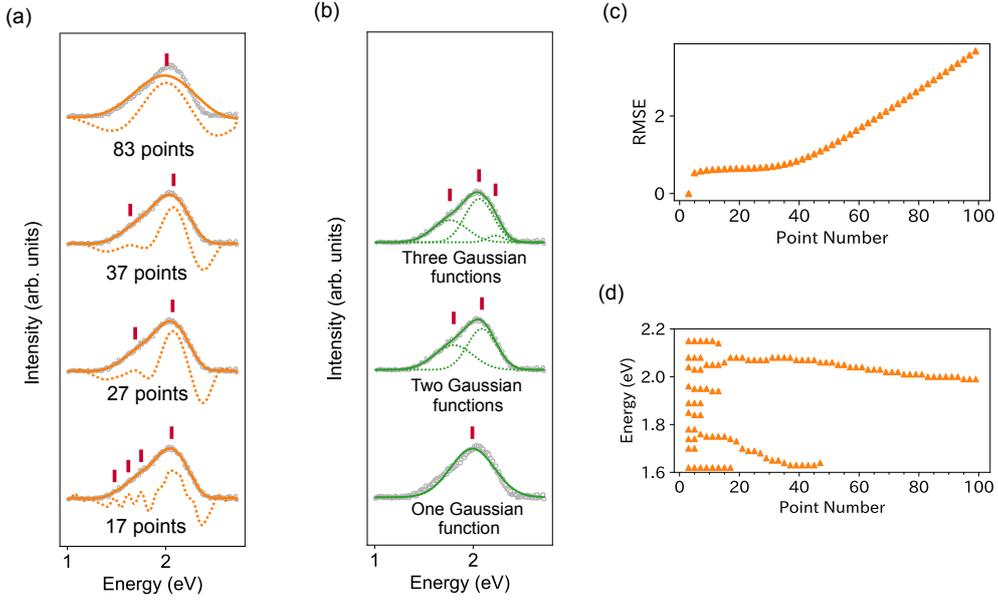

FIG. 3. Peak separation analyses of the experimental LEIPS spectrum (gray circles) using the second derivative and curve fitting methods. Red bars indicate the determined peak positions.
(a) Smoothed spectra using the SG method (orange solid lines) and their negative second derivative spectra (orange dashed lines). (b) Fitted spectra (green solid lines) and their component Gaussian functions (green dashed lines). (c) RMSE ($E_{RMS}^s$) as a function of the number of smoothing points. (d) Peak energies determined by the second derivative analysis.

Next, we analyze an experimental LEIPS spectrum to test the criteria established in Sec. IV A. The background signals of the LEIPS spectrum are approximated by a linear function and subtracted. First, we obtain the peak energies of the LEIPS spectrum through the second derivative analysis. Figure 3(a) shows the LEIPS spectrum, the smoothed spectra, and the negative second derivative spectra. At 17 smoothing points, many extra peaks appear. At 27 and 37 points, two peaks are apparent. Finally, only one peak is observed at 83 points. As in Sec. IV A, the number of smoothing points needs to be optimized. Figures 3(c) and (d) show the variation of $E_{RMS}^s$ and peak energy as a function of smoothing points. According to the criterion, the reasonable smoothing points are optimized to be between 19 and 35 points. In that region, two peaks are obtained at approximately 1.63–1.73 eV and 2.07–2.08 eV. The uncertainty of the peak at approximately 1.7 eV is estimated to be 0.10 eV.

We separate the spectral features by curve fitting using linear combinations of one to three Gaussian functions [Fig. 3(b)]. Whereas a single Gaussian function cannot correctly reproduce the experimental LEIPS spectrum, two and three Gaussian functions can reasonably match the spectral line shape. The peaks can be fitted by two Gaussian functions with similar widths, and their energies are 1.80 eV and 2.09 eV. Peak energies at 1.80 eV differ from those at 0.09 and 0.12 eV obtained by the deconvolution (discussed in Fig. 4) and the second derivative, respectively. When the number of Gaussian functions is three, the two Gaussian functions at peak energies of 1.76 and 2.06 are similar to the case of two Gaussian functions, and the intensity of the third Gaussian function at 2.27 eV is low. Therefore, we conclude that the spectral feature can be separated by two Gaussian functions. Generally, we do not know the number of Gaussian functions *a priori*. Furthermore, it is doubtful whether the peak shape is approximated by the Gaussian function. In fact, the line shape should be asymmetric owing to the interaction between the electron and the molecular vibrations (the Franck-Condon principle), as demonstrated in the HOMO-derived peak of an organic thin film[27].

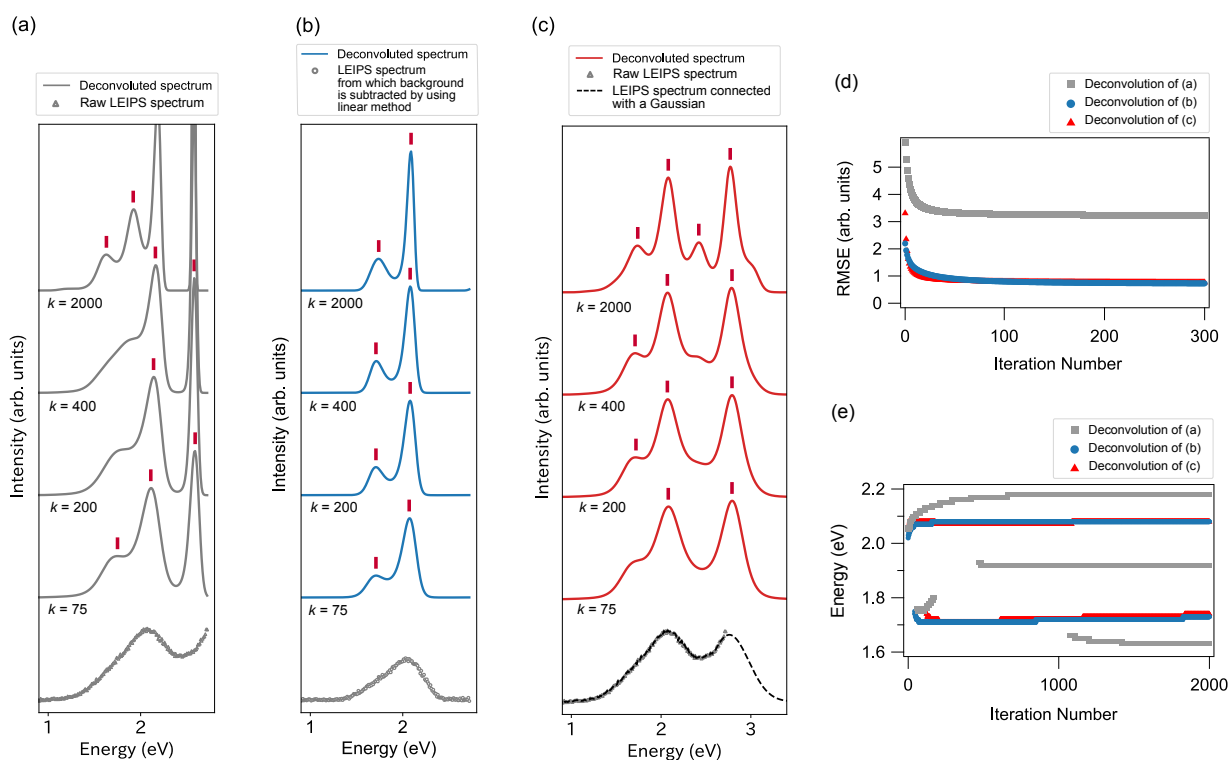

**FIG. 4. Deconvolution analysis of LEIPS spectra with different background treatments. (a) Raw LEIPS spectrum. (b) LEIPS spectrum with background subtracted using a linear method. (c) LEIPS spectrum connecting the high-energy edge to the baseline with a smoothed Gaussian. Red bars indicate the peak positions in the deconvoluted spectra. (d) RMSE ($E^{\text{d}}_{\text{RMS}}$) values of the deconvoluted spectra in (a), (b), and (c) are plotted as gray, blue, and red solid lines, respectively. (e) Peak energies determined from the deconvoluted spectra of (a), (b), and (c) are shown in gray, blue, and red solid plots, respectively.**

Finally, we discuss the deconvolution results. The results are significantly affected by the background signal. As shown in Fig. 4(a), the background signal rises with the increase in energy in the experimental LEIPS spectrum, whereas a constant background is assumed in the model spectrum. Figure 4(a) shows the deconvolution results for the raw LEIPS spectrum. At $k > 70$, a sharp peak appears at 2.59 eV in addition to the two broad peaks attributable to the LUMO bands at around 1.7 and 2.1 eV. We consider this sharp peak to be an artifact because this peak is never identified by the other methods and disappears depending on the background subtraction. When $k$ is increased, the peak at around 1.7 eV is obscured ($k = 200$ and 400) or the LUMO band splits into three peaks ($k = 2000$). Figure

4(d) shows the variation of $E_{RMS}^d$ as a function of $k$. Using the criteria in Sec. IV A, we judge that $E_{RMS}^d$ converges at around $k = 100$. Figure 4(e) shows peak energy dependence on $k$. Because the peak number changes from one to three in the region where $E_{RMS}^d$ converges ($k > 100$), the correct peak positions are not determined.

Figure 4(b) shows the deconvoluted LEIPS spectrum after subtracting the background signal approximated by a linear function. In this case, the sharp peak at 2.59 eV disappears, and two peaks are obtained. By observing the $E_{RMS}^d$ variation, we judge that the deconvolution calculation converges at around $k = 160$ [Fig. 4(d)]. As shown in Fig. 4(e), the peak energies of $\hat{o}^{160}(x)$ are at 1.71 and 2.07 eV. Even when the iteration number markedly increases [e.g., $\hat{o}^{2000}(x)$], the peak energies differ by only 0.02 eV compared to that of $\hat{o}^{160}(x)$. Therefore, peak energy dependence on iteration number is sufficiently small, and background subtraction is satisfactory.

The above analysis indicates that background subtraction is critical and that the abrupt change in the high-energy end of the raw LEIPS spectrum may result in an artificial peak. Therefore, we smoothly connect that spectral edge to the baseline through a Gaussian function, as shown in Fig. 4(c). Through deconvolution, the peak at around 2 eV is obtained at $k = 75$. The two peaks are obtained at approximately 1.7 and 2.1 eV for $k = 200, 400$, and 2000.

Using the criteria in Sec. IV A, $E_{RMS}^d$ converges at approximately $k = 120$ [Fig. 4(d)]. The peak energies of $\hat{o}^{125}(x)$ are 1.74 eV and 2.08 eV, in agreement with that obtained by linear background subtraction. From Fig. 4(e), the uncertainty of peak energy is only 0.03 eV from $k = 125$ to an excessive iteration (e.g., $k = 2000$). This background processing is also satisfactory. To summarize the deconvolution results, we find that the deconvoluted spectrum is distorted if the background is not properly processed. The distortion can be avoided by subtracting the background or by connecting the spectral edge to the spectral baseline using a smooth function. The background of the LEIPS spectra is subtracted using the linear method hereinafter, as this procedure is widely used in spectral analysis.

We summarize the peak separation of the model and the experimental LEIPS spectrum. In the second derivative, the results are influenced by the number of smoothing points. Insufficient smoothing can introduce additional peaks due to noise. On the other hand, excessive smoothing distorts the line shape of the peaks and alters their energy. The curve fitting method has difficulty finding reasonable model functions, namely, the type and number of functions. In this study, we find that the two Gaussian functions yield reasonable results. However, it is doubtful whether the Gaussian function is a good model for representing the LEIPS spectral line shape. In the deconvolution, the results are influenced by the termination of the iteration number and the background processing. The optimal iteration number can be determined from the variation of $E_{RMS}^d$. For the background processing, it is essential to ensure that both ends of the spectrum are smoothly connected to the baseline, which can be achieved by subtracting the background or connecting a smooth curve to the end of the spectrum.

# V. Discussion

## A. Robustness of the three methods to noise

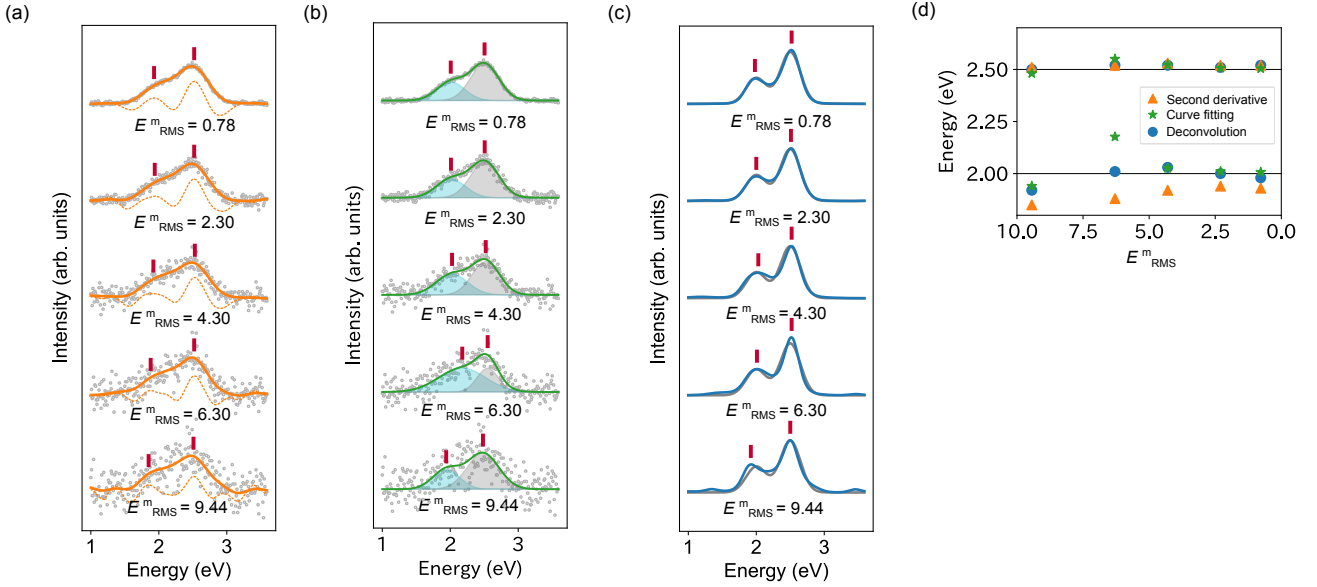

**FIG. 5. Analysis of model spectra (gray circles) with varying noise levels using the three methods. Noise level is defined as RMSE of the model spectra ($E_{RMS}^m$). Red bars indicate the determined peak positions. (a) Second derivative analysis. Smoothed spectra using the SG method (orange solid lines), and their negative second derivative spectra (orange dashed lines). (b) Curve fitting analysis. Fitted spectra (green solid lines) and their component Gaussian functions (filled areas) (c) Deconvolution analysis. Deconvoluted spectra (blue solid lines) and original spectra (black solid lines). (d) Peak energies determined by deconvolution (blue), second derivative (orange), and curve fitting (green). Black solid lines represent the true peak energies.**

In this section, we discuss the effects of noise on the results of peak separation. First, the model spectra with different noise quantities are analyzed. The procedure for preparing the model spectra is described in Sec. III. The results of second derivative, curve fitting, and deconvolution analyses are shown in Figs. 5(a), 5(b), and 5(c), respectively. The peak energies are summarized in Fig. 5(d). The true peak energies of the original spectra (2 eV and 2.5 eV) are shown by black solid lines. The three methods can correctly determine the peak energy of Gaussian 2 within an energy error of 0.05 eV regardless of $E_{RMS}^m$.

In contrast, the obtained peak energy of Gaussian 1 differs depending on the methods. In the second derivative analysis, the obtained peak energy of Gaussian 1 differs by 0.12 eV from the true peak energy when $E_{RMS}^m$ is 9.44. As $E_{RMS}^m$ decreases, the agreement is improved, and the difference is 0.07 eV at $E_{RMS}^m = 0.78$. In the curve fitting analysis, the width and intensity ratio of the fitted Gaussian functions are inconsistent with Gaussians 1 and 2 at $E_{RMS}^m = 6.30$. At $E_{RMS}^m \leq 4.30$, they are reasonable, meaning that the peak fitting is successful. The determined peak energy of Gaussian 1 agrees with the true peak energy within an error of 0.03 eV at $E_{RMS}^m \leq 4.30$. In the deconvolution, the deconvoluted spectrum is largely distorted at $E_{RMS}^m = 9.44$. The peak energy of Gaussian 1 differs by 0.08 eV from the true peak energy. In $E_{RMS}^m \leq 6.30$, the peak energies of Gaussian 1 correspond to the true peak energy within an error of 0.03 eV. These results indicate that curve fitting and deconvolution are more robust to noise than the second derivative.

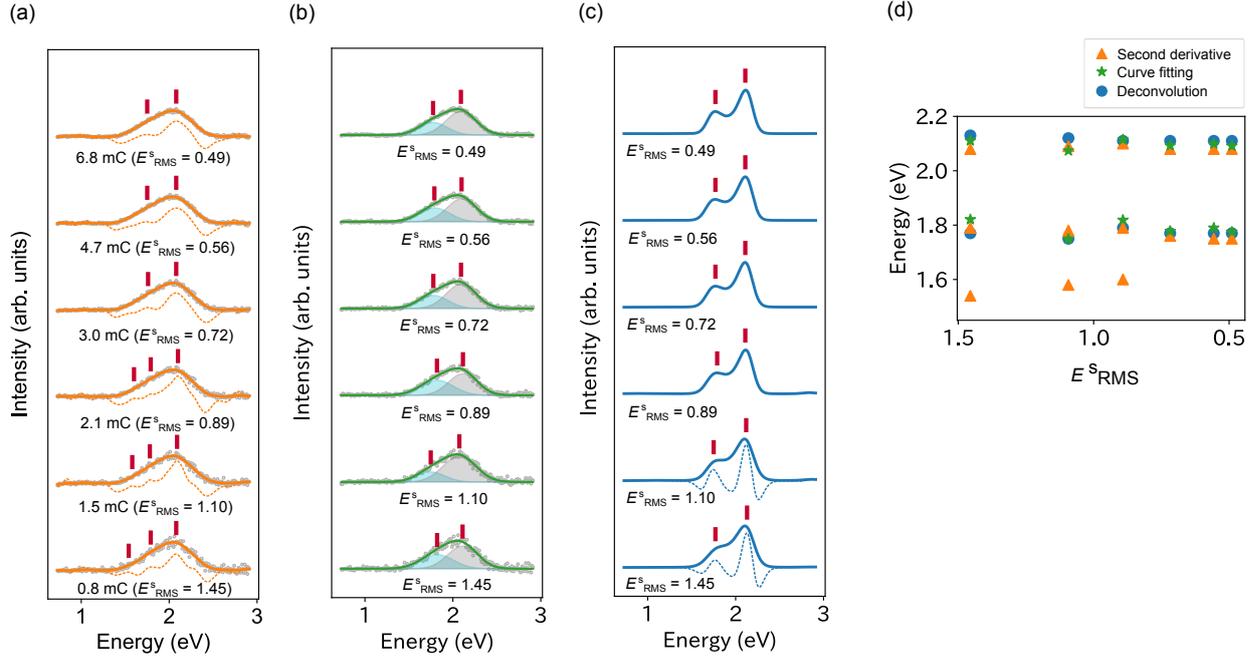

FIG. 6. Analysis of LEIPS spectra (gray circles) with varying total charge of the irradiated electron beam (equivalent to the measurement integration time), using the three methods. Red bars indicate obtained peak positions. (a) Second derivative analysis. Smoothed spectra (orange solid lines) and negative second derivative spectra (orange dashed lines). (b) Curve fitting analysis. Fitted spectra (green solid lines) and their component Gaussian functions (filled areas). (c) Deconvolution analysis. Deconvoluted spectra (blue solid lines) and their negative second derivative spectra (blue dashed lines). (d) Peak energies obtained by second derivative (orange), curve fitting (green), and deconvolution (blue).

Next, we examine the effect of noise on the peak separation in the experimental LEIPS spectra. Experimental details are described in Sec. III. The results of the second derivative, curve fitting, and deconvolution analyses are shown in Figs. 6(a), (b), and (c), respectively. The peak energies obtained by these analyses are summarized in Fig. 6(d) as a function of noise quantity expressed in terms of $E_{RMS}^s$. In the second derivative analysis, an additional peak appears at approximately 1.6 eV owing to noise at $E_{RMS}^s \geq 0.89$. At $E_{RMS}^s \leq 0.72$, the two peaks appear at around 1.79–1.75 eV and 2.08–2.10 eV. For the curve fitting analysis, the peak energies are apparent at approximately 1.75–1.81 and 2.07–2.11 eV. In the deconvolution, because the deconvoluted spectra at $E_{RMS}^s \geq 1.1$ are broad, the peak energies are determined by their second derivatives. The deconvolution yields two peaks at 1.75–1.78 and 2.11–2.13 eV regardless of $E_{RMS}^s$.

From the results shown in Figs. 5 and 6, we conclude that deconvolution and curve fitting are more robust to noise than the second derivative. The second derivative, curve fitting, and deconvolution analyses can obtain reasonable peak energies at $E_{RMS} \leq 0.72$, $E_{RMS} \leq 4.3$, and $E_{RMS} \leq 6.3$, respectively.

## B. Peak separation capability

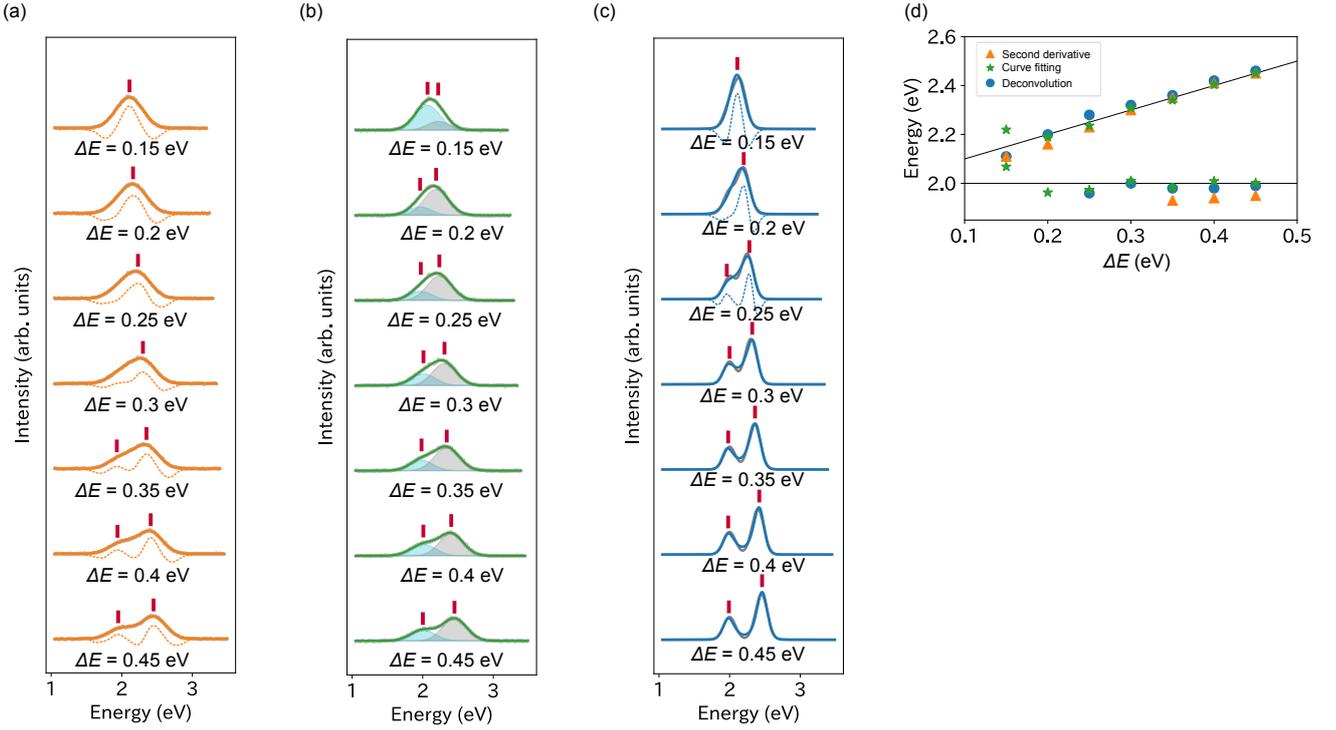

FIG. 7. Analysis of model spectra (gray circles) with varying peak energy differences ΔE between Gaussians 1 and 2 using the three methods. Red bars indicate the obtained peak positions.
(a) Second derivative analysis. Smoothed spectra (orange solid lines) and negative second derivative spectra (orange dashed lines). (b) Curve fitting analysis. Fitted spectra (green solid lines) and component Gaussian functions (filled areas). (c) Deconvolution analysis. Deconvoluted spectra (blue solid lines), negative second derivative spectra (blue dashed lines), and the original spectra (gray solid lines). (d) Peak energy determined by second derivative (orange), curve fitting (green), and deconvolution (blue). The true peak positions of Gaussians 1 and 2 are shown as black solid lines.

In this section, we discuss the peak separation capability by analyzing the model spectra $m(x)$. Generally, the positions of two peaks with an energy difference smaller than the peak widths are challenging to identify. The procedure for preparing model spectra is described in Sec. III. The results of the second derivative, curve fitting, and deconvolution analyses are shown in Figs. 7(a), (b), and (c), respectively. Figure 7(d) summarizes the obtained peak energies as a function of energy spacing ΔE of Gaussians 1 and 2. The second derivative separates peaks in $m(x)$ with ΔE ≥ 0.35 eV. The curve fitting determines the peak energies of $m(x)$ with ΔE ≥ 0.2 eV. In the deconvolution, the peak energies are determined by their second derivative in the case of $m(x)$ with ΔE ≤ 0.25 eV. The deconvolution determines the peak energy of $m(x)$ with ΔE ≥ 0.25 eV.

We conclude that the peak separation capability of deconvolution and curve fitting is superior to that of the second derivative. Note that this conclusion depends on the relationship among the FWHMs of Gaussian 1, Gaussian 2, and instrumental function $i(x)$. When the FWHM of Gaussians 1 and 2 is 0.35 eV, the peak separation capability of the three methods is almost the same (see Fig. S9 in Supplementary Material). The three methods reliably separate the peak energies when ΔE exceeds the FWHM of $i(x)$. Because the deconvolution only removes the broadening due to

*i*(*x*) from *m*(*x*), its advantage over the second derivative is slight if the FWHM of *o*(*x*) is greater than that of *i*(*x*).

# VI.   *Conclusion*

To determine the positions of the two peaks hindered by experimental broadening in the LEIPS spectrum, we compared three peak separation methods—second derivative, curve fitting, and deconvolution. We analyzed two closely spaced peaks, taking into account the LUMO band of pentacene. We analyzed the model spectrum composed of two Gaussian functions and the experimental LEIPS spectrum.

In the second derivative analysis, the energy of the overlapping peaks was obtained from the peaks of the negative second derivative spectrum. The measured spectrum must be smoothed before the iteration. The results were influenced by the number of smoothing points. Insufficient smoothing introduced artificial peaks, whereas excessive smoothing altered the energy of the detected peaks and caused peak structures to disappear. There was no reliable method to determine the number of smoothing points. Additionally, the second derivative was easily affected by noise, requiring a high S/N ratio of the experimental spectrum. Conversely, the second derivative did not require background subtraction, facilitating data analysis when the exact background shape was unknown.

In curve fitting, the assumed fitting function was a critical condition. It was essential to choose a physically plausible function and the number of functions. In the case of LEIPS (and also UPS) spectra, where the Frank-Condon principle governed the peak shape, selecting the appropriate function was not straightforward. Furthermore, the background subtraction was necessary before the fitting procedure. Curve fitting was more robust to noise than the second derivative. This method was powerful when the fitting function was already known.

In the deconvolution analysis, Jansson's method successfully restored the original spectrum. The main deconvolution parameter was the termination of the iteration number, which could be determined from the convergence of $E_{\text{RMS}}^{\text{d}}$. The discontinuity of the spectrum severely distorted the deconvolution result. To avoid this distortion, subtracting the background or linking the spectral edge to the spectral baseline was necessary. The deconvolution and curve fitting methods could separate more closely spaced peaks than the second derivative method.

Thus far, both second derivative and curve fitting methods have been frequently used to analyze PES spectra. However, LEIPS spectra are broadened by the instrumental function, making deconvolution a valuable and practical method as well.

## Supplementary materials

See supplementary materials for additional information on the second derivative analysis of a model spectrum, the deconvolution analysis of a model spectrum, and a discussion on peak separation capability.

## Conflict of interest

The authors have no conflicts of interest to disclose.

## Ethics approval

Ethics approval not required.

# Author contributions

**Ryotaro Nakazawa:** Conceptualization (equal); Data Curation (lead); Formal Analysis (lead); Methodology (lead); Project Administration (lead); Software (lead); Supervision (equal); Validation (equal); Visualization (lead); Writing/Original Draft Preparation (lead); Writing/Review & Editing (equal)

**Haruki Sato:** Formal Analysis (equal); Investigation (lead); Writing/Review & Editing (equal)

**Hiroyuki Yoshida:** Conceptualization (equal); Methodology (equal); Project Administration (equal); Resources (lead); Supervision (equal); Writing/Review & Editing (lead)

# Data availability statement

Supporting data for this study are available from the corresponding authors upon reasonable request.